# Development and preclinical test of a $^{32}$P containing PEEK foil aiming at urethral stricture prevention by LDR brachytherapy

Walter Assmann[1], Ricarda Becker[2], Christian Stief[3], Ronald Sroka[2,3]

[1] Department of Medical Physics, Ludwig-Maximilians-Universität München, Garching, Germany
[2] Laser-Forschungslabor, LIFE Center, University Hospital, Ludwig-Maximilians-Universität München, München, Germany
[3] Department of Urology, University Hospital, Ludwig-Maximilians-Universität München, München, Germany

## Abstract

*Purpose.* Primary or recurrent urethral stenosis are a common clinical problem. The aim of this feasibility study was development and application test of a novel radioactive catheter for potential use in LDR brachytherapy to prevent urethral stenosis. *Methods.* A beta radiation emitting $^{32}$P-foil was developed and integrated in an indwelling catheter, which is usually inserted after surgical interventions in the urethra. Activity and dose distribution were measured after neutron activation of the $^{31}$P foil component with scintillation techniques as well as radiochromic films and compared to MC simulations. Practicability and radiation safety of this new applicator was tested on male rabbits, which underwent before catheter application a new method of stricture induction by radial laser light irradiation. *Results.* Measured dose distributions of catheters with $^{32}$P-foils were found in good agreement with MC simulations. Wash-out test confirmed the radioactive catheter setup to be inside the permitted limits of a sealed source. Strictures could be induced by laser light in all animals, surgically treated by slitting and irradiation with a 7-day dose of 15 Gy or 30 Gy without adverse radiation effects during 4 weeks observation time. *Conclusions.* This prove-of-concept study presents a save and technically uncomplicated approach of LDR brachytherapy intending to prevent urethral stenosis. The prescribed irradiation dose can be administered in a reproducible and precise manner using a urethral catheter as carrier of a beta radiating foil. The study duration of 63 days was too short for statistically significant stricture related conclusions. This approach may be considered for similar problems in other hallow organs.

## Introduction

Brachytherapy is usually associated with treatment of malignant diseases, and the most common example is local radiotherapy of the prostate tumor using $^{125}$I seeds. Less well known are the possibilities of brachytherapy in treatment of benign diseases, in which radiotherapy can also refer to decades of positive experience (Seegenschmiedt *et al* 2008, Seegenschmiedt *et al* 2015). Examples are prevention of heterotopic ossification after hip arthroplasty and inhibition of disease progression in different (hyper)proliferative or inflammatory disorders (Trott 1994). Another field of application is prevention of in-stent restenosis by intravascular brachytherapy (Khattab *et al* 2021), for coronary (Fox 2002) as well as peripheral arterial occlusion diseases (Hansrani *et al* 2002). The radio-biological mechanisms are still not fully enlightened, but the formation of both inflammatory cytokines and those of certain growth factors can be reduced by ionizing radiation with doses in the range of 10 to 20 Gy (Rubin *et al* 1998, Arenas *et al* 2012). Since in benign diseases unnecessary radiation exposure of healthy tissue is even less tolerated than in case of tumor irradiation, brachytherapy with low irradiation depth and thus reduced damage to the surrounding tissue of the target volume is in advantage over teletherapy. This makes $^{32}$P a radioisotope of choice for applications where the radioactive source is in intimate contact with the targeted tissue. $^{32}$P emits electrons of 690 keV mean energy (1.7 MeV maximal energy), leading to a maximum range in tissue of about 6 mm.

A particularly suitable field of $^{32}$P application for LDR brachytherapy could be prevention of (re-)stenosis in hollow organs, if after surgical intervention excessive scarring during wound healing occurs requiring further medical intervention. Wound healing begins with an inflammatory phase, which dominates in the first two days after the injury, followed by a proliferation phase in the following days. Both processes can be influenced by ionizing irradiation, and $^{32}$P with half-life of 14.3 days fits well this time scale of wound healing. Successful treatment of hyperproliferative disorders with specially developed $^{32}$P implants has been demonstrated in preclinical studies on glaucoma filtering surgery (Assmann *et al* 2007) or prevention of restenosis of paranasal neo-ostia (Oestreicher *et al* 2017). However, especially for LDR brachytherapy, there is only a very limited number of approved radioactive implants available, which restricts the possibilities for brachytherapy applications.

In this report, development of a new kind of radioactive $^{32}$P-foil will be presented, intended for adaption to very different application geometries, but for use in hollow organs in particular. This concept and its feasibility were tested in a dedicated preclinical in-vivo model setting associated with prevention of benign urethral stenosis. Primary radiation effects were observed, the study duration, however, excluded further radiobiological investigations on stenosis development.

## Materials and methods

The radioisotope $^{32}$P is usually generated by activation of stable $^{31}$P via the neutron capture reaction $^{31}$P(n, γ)$^{32}$P, which has a capture cross-section for thermal neutrons of 170 mbarn only, the activation is therefore best carried out at a high-flux neutron source. $^{31}$P activation can be performed before or after incorporation into a therapeutic applicator. The first possibility, originally developed for $^{32}$P coronary stent fabrication by ion implantation (Huttel et al 2002, Heesch et al 2018), was already applied in pre-clinical studies for neo-ostium stents (Oestreicher *et al* 2017) and for bioresorbable implants in glaucoma therapy (Assmann *et al* 2007). Although ion implantation has shown to be quite flexible, nevertheless, this technique was considered to be too complex and costly. Therefore, an easier production process was chosen starting from $^{31}$P containing foils, which can be prepared and stored for later activation on demand.

*Development of $^{31}$P-containing foils and prototype production*

Due to the neutron activation process, the foil polymer has to withstand a neutron and gamma dose of many MGy without significant changes of material properties. Extensive irradiation tests on several biocompatible polymers showed, that polyether ether ketone (PEEK) meets these requirements best. This polymer is proved to be biocompatible, chemically very resistant, and applicable up to at least 250 °C. PEEK melts compared to most other thermoplastics at a relatively high temperature of around 341 °C. In the range of its melting temperature it can be processed using cost-effective extrusion methods of granulate mixed with a $^{31}$P containing additive. A suited additive has to be stable up to the PEEK melting temperature and should not contain components forming long-lived radioactive nuclei during neutron activation. Sodium metaphosphate (NaPO$_3$) was selected as phosphoric compound because it is nontoxic, available in powder form and thermally stable up to 550 °C. Besides $^{32}$P only the short-lived gamma emitter $^{24}$Na (T$_{1/2}$ = 15h) is created during activation, which can even serve as an additional monitor to cross-check the $^{32}$P activity. The foil production started by mixing of PEEK granules (PEEK 381G natural, Victrex, England) with 25 wt% NaPO$_3$ powder in a screw compounder and subsequent melt filtration to get a fine-particle compound granulate. From this material, a 50 μm thick foil was extruded, which showed a homogeneous NaPO$_3$ distribution in first transmitted light inspection. Due to the use of an experimental production plant the variation of the NaPO$_3$ content was to be expected about 15%, therefore the exact final $^{32}$P activity of an individual foil implant must be determined later.

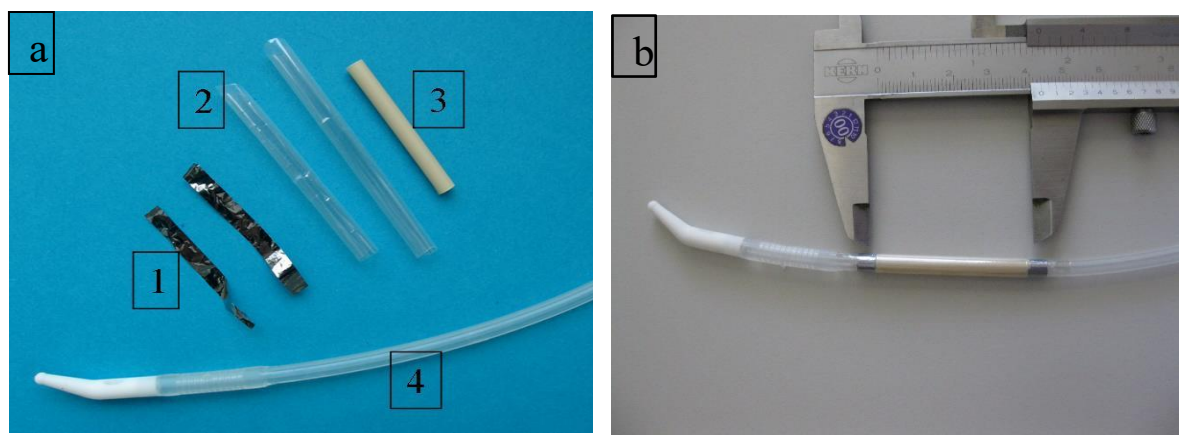

**Fig. 1:** $^{32}$P-foil catheter prototype. (a) components: (1) Ta X-ray marker foil, (2) shrink tubing, (3) $^{32}$P-foil roll (4) urethral catheter; (b) prototype fully assembled.

The novel approach of this study was to use a urethral catheter as carrier of the activated $^{31}$P-foil, as after surgical interventions in the urethra an indwelling catheter is inserted anyway. This approach has the advantage that the later clinical application procedure of the catheter could remain unchanged except for the necessary radiation protection measures. Particular attention was drawn to a sufficient length of the foil, so that the stenosis area could be irradiated with the desired dose even with possible positioning errors and displacements of the implant, and thus to avoid the so-called edge effect known from coronary $^{32}$P-stents (Albiero 2000b). The observed stent end stenosis had been traced back at that time to the dose drop at the stent ends, where still injured areas heal in the presence of low radiation levels, which can even stimulate proliferation. Hence, rectangular pieces were cut from the $^{31}$P-foil stock of 4 cm length and around 3 cm width fitting to form a double-layer around the catheter. These two foil layers may reduce dose inhomogeneities due to eventually unequal phosphor grain distribution, and on the other hand halve the neutron activation time. The roll reached thermoplastic form stability by tempering at 220 °C for two hours. Additionally, a 10 mm diameter circular plate was punched out from the same foil stock area, which served as monitor for quality assurance. Foil rolls and monitor plates were weighted and only those further processed which were within a 5% area weight window, i.e. with corresponding NaPO$_3$ content. Due to the production process NaPO$_3$ is distributed within the foil up to the surface and could potentially be solved and washed out during later application. Therefore, before activation, all foil rolls were thoroughly washed in a shaker for two hours in a mixture of 90% VE water and 10% purest ethanol. Selected pieces were welded under argon atmosphere in dedicated activation capsules. For final assembly of the urethral catheter (indwelling catheter, Rüsch-Care, Teleflex Medical GmbH, Germany), after foil activation and an appropriate decay time for $^{24}$Na, a 10 mm thick plexiglass shielding was used for radiation protection. The $^{32}$P-foil roll was slid on the catheter tube and two Ta X-ray markers were glued to both foil ends (Fig. 1). Various sealing concepts were tested to prevent later $^{32}$P wash-out, either one or two overlapping biocompatible shrink tubings (Polyester, Advanced Polymers, Salem NH, USA). Alternatively, one shrink tube with circular end sealing was tried by means of different adhesives such as Loctite 4061 (Medical Line, Loctite, Garching, Germany), Technomelt (Henkel, Munich, Germany), and Dymax 222/100 (Dymax Europe, Frankfurt, Germany). Before application in animal tests, the whole foil-catheter setup was sterilized by 30 kGy $^{60}$Co gamma radiation.

*Activity and dose distribution determination*

Neutron activation was performed by using the capsule irradiation system of the Research Neutron Source (FRM II) of the Technical University of Munich (TUM, Garching, Germany), which offers a thermal neutron flux density of up to $10^{14}$ cm$^{-2}$ s$^{-1}$. Due to the position dependent neutron flux each capsule contained an additional Au standard for determination of the particular integrated neutron dose by gamma spectroscopy. After neutron activation of 20 minutes typical duration, implant prototypes were first subjected to measurements of $^{32}$P activity including wash-out tests and determination of dose distribution, which required special effort due to the low range of electrons of only a few millimeters. Dose measurements were performed according to existing protocols of the AAPM TG-43 and TG-60 for interstitial brachytherapy sources with low-energy gamma radiation (Nath 1995, Nath 1999). For direct determination of the $^{32}$P activity a liquid scintillation counter (LSC) (Tri-Carb 2900 TR with scintillator Ultima Gold AB, Perkin Elmer, Germany) was used. This method, however, is calibrated only for liquid radioactive materials, whereas PEEK has a very poor solubility. The unknown detection efficiency for activated monitor plates was thus calibrated via gamma spectroscopy of the stoichiometric component $^{24}$Na, which decays to almost 100% by emission of high energetic gamma radiation (1368.6 keV, 2754.0 keV). Potential $^{32}$P wash-out of the radioactive catheter setup was also determined by means of LSC.

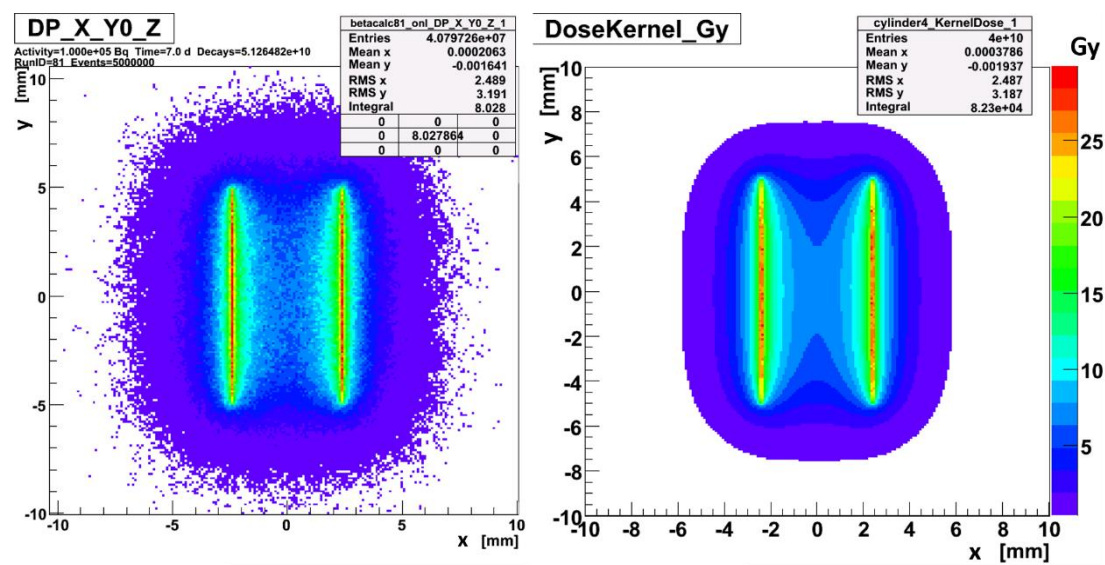

**Fig. 2:** 7-day dose distribution of a $^{32}$P-foil roll in water with 1 cm foil length, 14 Fr diameter and 100 kBq start activity, simulation by Geant4 (left) and KernelCalc (right) shown in a plane through the roll axis.

Dose distributions of $^{32}$P-foil rolls were both calculated and measured for water environment as proposed in Caswell (2004), in view of the later close contact to urethral tissue. MC simulation toolkit *Geant4* (Agostinelli 2003) was used to model the beta decay of $^{32}$P and the related electron dose deposition including Bremsstrahlung, which tracks electrons down to an energy threshold of 150 eV. An example of a Geant4 dose simulation of a 14 Fr catheter equipped with a $^{32}$P-foil roll of 1 cm length and 100 kBq activity is shown in Fig. 2 (left), which demonstrates the short range of the emitted beta radiation and the therewith easy radiation protection by 10 mm thick plexiglass. Additionally, a much faster dose point kernel calculation was developed ("KernelCalc") using a superposition of pre-calculated, equally distributed point-like $^{32}$P sources, an example is given in Fig. 2 (right) for comparison to Geant4. Dose distributions calculated with both methods were found in good agreement as shown in Fig. 3, where the radial Geant4 dose distribution is compared to that of the dose kernel calculation. The intended tissue dose of 15 Gy at 1 mm distance from the foil surface can be reached within an exposure time of 7 days by 264 kBq start activity only. It should be noted, this cumulated 15 Gy dose results in a tissue contact dose of about 80 Gy.

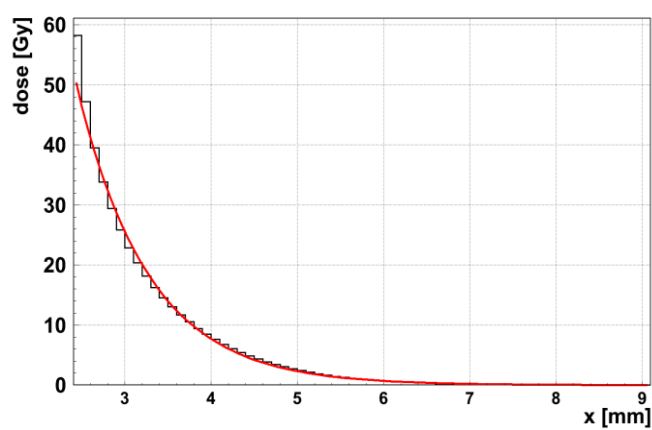

**Fig. 3:** Radial dose distribution of the $^{32}$P-foil roll shown in Fig. 2 at y = 0 mm: simulation by Geant4 (black) and KernelCalc (red), dose from foil surface outwards.

For experimental verification of the calculated dose distributions measuring techniques are required with about 1-mm spatial resolution. The OPTIDOS system (PTW, Freiburg, Germany) fulfills this demand, which was developed especially for brachytherapy sources such as prostate seeds. The actual detector consists of a cylindrical scintillation crystal with 1 mm diameter and 1 mm length, thus achieving the required spatial resolution in the range of 1 mm and absolute accuracy within an error of 16%. The dose distribution for source distances below 2 mm were measured with a radiochromic film stack (GafChromic EBT, ISP, Wayne NJ, USA), whose single film thickness of about 0.3 mm provides an adequate depth resolution. The EBT films were calibrated with 6 MV photons using the known dose equivalence to electrons within an error of 7% (Richter 2009).

*Animal model and preparatory experiments*

Male rabbits (New Zealand line; aged 6–7 months; weight: 3.5–4 kg) were selected as in-vivo model, as their urethra can be treated with existing instruments from the pediatric urology. Three treatment-groups (each n = 6 animals) with 30 Gy or 15 Gy cumulated 7-day dose in 1 mm urethra tissue depth and 0 Gy (control) were enrolled. The targeted dose with the $^{32}$P implant was based on the studies for prevention of coronary in-stent restenosis with $^{32}$P stents (Albiero 2000a), since the urethra of rabbits has comparable dimensions. The following study protocol was developed for a small cohort in-vivo feasibility study with specific regard to a proof-of-concept application on urethral stricture guided by the clinical procedure: Induction of a urethral stricture (day 0); check and surgically opening of the stricture by urethrotomy (according to Sachse), insertion of catheter (day 28); removal of catheter after 30, 15 or 0 Gy cumulated dose (day 35); rabbit sacrifice, urethra removal and preparation for subsequent histological analysis including different staining procedures (e.g.: H&E, EvG) (day 63). At each step the urethra will be characterized by an optical and a radiological method (videourethroscopy and cystourethrography). This study protocol was in accordance with the German Animal Welfare Act and approved by the local authorities of Upper Bavaria (approval no 55.2-1-54-2531-69-08).

Two preparatory experiments were performed to test new methods for positioning control of the urethral catheter on the one hand and for induction of a urethral stenosis on the other hand. Due to the steep dose drop of beta radiation a close contact of the urethral tissue and the $^{32}$P-foil was essential, which depends mainly on a proper size of the catheter. This tissue-foil contact was checked with optical coherence tomography (OCT) on two rabbits. Endoluminal radial OCT investigation was performed using a time-domain OCT (Type M2x, LightLab Imaging Inc., UK) with light of a superluminescence diode (1300 nm, 25 mW) (Püls2009). After insertion of a 14 Fr catheter into the bulbar urethra the radial OCT probe was introduced into the central passageway lumen of the catheter. Within the relevant range, the tissue was in close contact to the implant up to a few places of a few millimeters in length, where a gap of a maximum of 0.4 mm occurred (Fig.4).

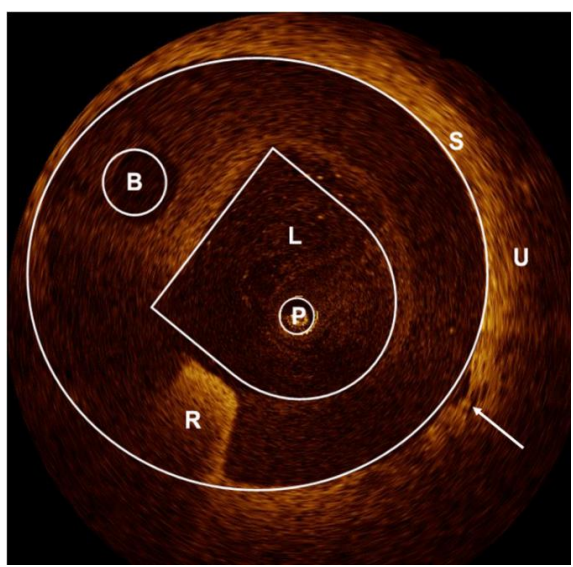

**Fig. 4:** OCT-cross section image of a 14 Fr catheter inside a rabbit urethra: (P) radial OCT probe, (L) passageway lumen, (B) balloon dilation channel, (R) reflex of radiopaque material, (S) outer catheter surface, (U) urethral tissue. The arrow points to a 0.4 mm gap between catheter and tissue.

Urethral strictures were induced by a technique, which was developed originally for endovenous laser ablation of varicose veins, using laser light (1470 nm diode laser) via an endoscopically introduced, radially emitting laser fiber (Sroka 2015). Precise positioning of the fiber and reproducible radial irradiation were enabled by a circular pilot light under optical control. Application of about 100 J light energy was found sufficient for successful circumferential thermal destruction of the urethral tissue.

# Results

Strictures of the urethra represent a relevant problem among urological patients and are characterized by increased tissue formation and scar contraction during the wound healing process following surgical interventions. The aim of this feasibility study was primarily focused on tests of our novel foil implant: their suitability for stricture prevention, their practicality in a clinical environment and their safety of application with regard to radiation protection.

*Measurements on activated prototypes*

Simulated and calculated dose distributions were first checked on monitor plates, where both dose and activity can be measured easily and extrapolated by simulations to the corresponding foil rolls. A dose distribution, measured by an EBT film stack and corrected for the water-equivalent thickness, is compared in Fig. 5 to a KernelCalc calculation using the LSC determined plate activity as input. Additionally, in Fig. 5 at 2 mm distance, the dose value measured with the calibrated OPTIDOS system is plotted confirming the overall good agreement of the different methods within the estimated experimental error of 7% (EBT) and 16% (OPTIDOS), respectively. The lateral EBT resolution offered also a possibility to control the $^{32}$P grain distribution, which was found to be sufficiently homogeneous without "hotspots", and will be additionally smoothed by two foil layers. Measuring both $^{24}$Na activity by gamma spectroscopy and $^{32}$P activity with LSC of monitor plates a linear relation was found between activity and weight, therefore the intended implant activity could be adjusted to 5% precision by selecting foils of a 5% surface weight range.

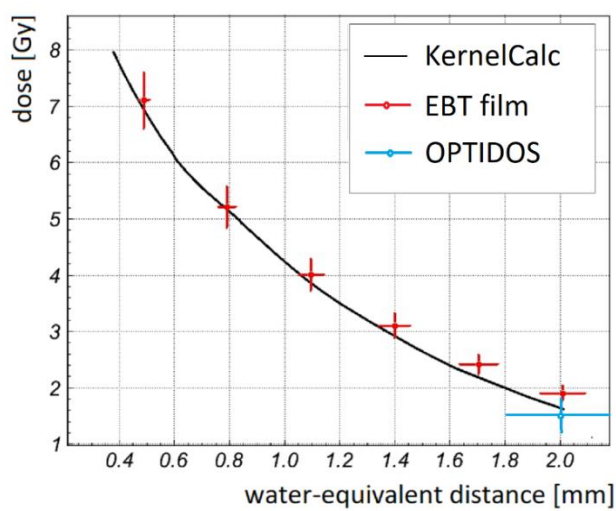

**Fig. 5:** Dose distribution of a $^{32}$P monitor foil plate in water (measurement with EBT film stack and OPTIDOS) compared to KernelCalc simulation for 35.8 kBq activity (as measured with LSC).

$^{32}$P wash-out tests of activated catheter assemblies with various sealing techniques yielded by far the best result for two overlapping shrink tubings. A cumulative wash-out for 8 days of only 0.02% of the start activity was measured (Fig. 6). A maximum of 200 Bq activity wash-out within 4 hours is allowed for a sealed radioactive source according to German standard DIN 25426-4 (DIN 2008), which can be easily fulfilled by a two-tubing sealing for at least 1 MBq start activity, four times more than the envisaged implant activity. Apparently, all

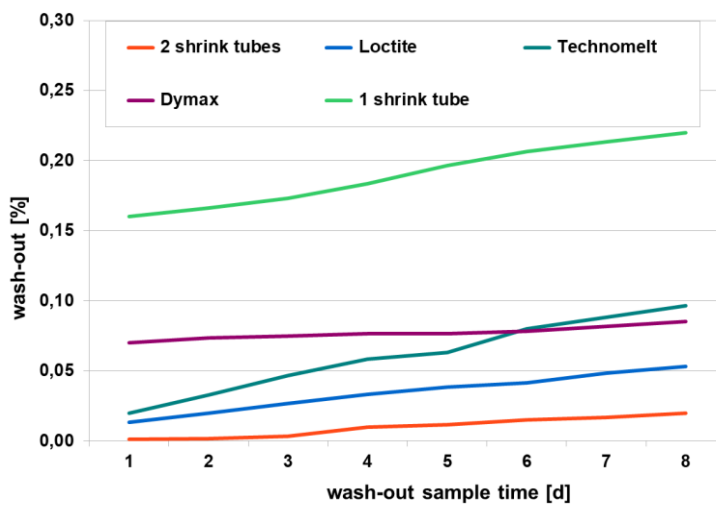

**Fig. 6:** Result of wash out tests using different sealing concepts with one or two shrink tubings without or with various end sealing adhesives (for explanation see text).

adhesives could not be securely enough attached at single shrink tube ends to prevent ingress of liquids by capillary effect and successive wash-out. Thus, $^{32}$P foils were finally sealed by means of two overlapping heat shrink tubes of 25 μm wall thickness as shown in Fig. 1.

*Application test of a urological $^{32}$P catheter prototype on rabbits*

First, according to the study protocol, a defined stricture had to be generated in the rabbit urethra located about 15 mm distal to the colliculus seminalis. After application of 100 J laser light energy a pronounced induced stricture developed during the following 28 days, which was characterized both by cystourethrography and videourethroscopy. According to the stricture classification as shown in Fig. 7, different stricture grades were observed: grade III in 15/18 animals (83%), grade II in 1/18 animals (6%) and grade I in 2/18 animals (11%). To mimic the clinical treatment procedure the strictures were surgically slit lengthwise at 12 o'clock until the full width of the lumen was reached. Immediately after surgical opening of the urethra, the gamma-sterilized 14 Fr catheter carrying the special developed foil roll was inserted. The foil position was matched with the stricture by radiography making use of the two X-ray markers, the catheter balloon was filled with contrast agent and blocked in the bladder neck. During the whole procedure beta radiation was shielded by a 10 mm thick plexiglass cylinder covering the $^{32}$P-foil area (Fig. 8). All people working with the radioactive catheter were equipped with ring batches and extremity dosimeter.

| Stenosis | X-ray | video | Images ||
|---|---|---|---|---|
| Grade I | Residual lumen > 50% | passable with the cystoscope | 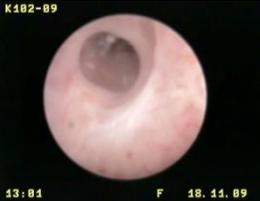 | 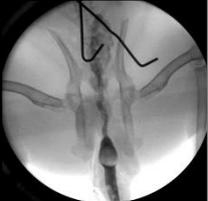 |
| Grade II | Residual lumen < 50% | only passable with the cystoscope after dilatation | 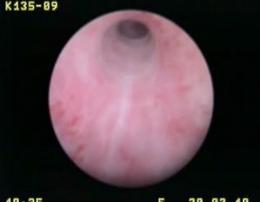 | 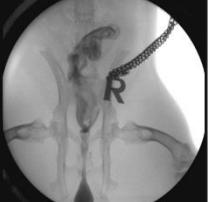 |
| Grade III | Residual lumen < 10% | not passable with the cystoscope | 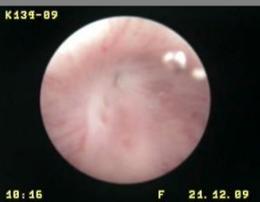 | 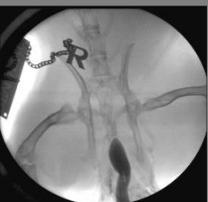 |

**Fig. 7:** Stenosis grade definition by cystourethrography (X-ray) and videourethroscopy (video).

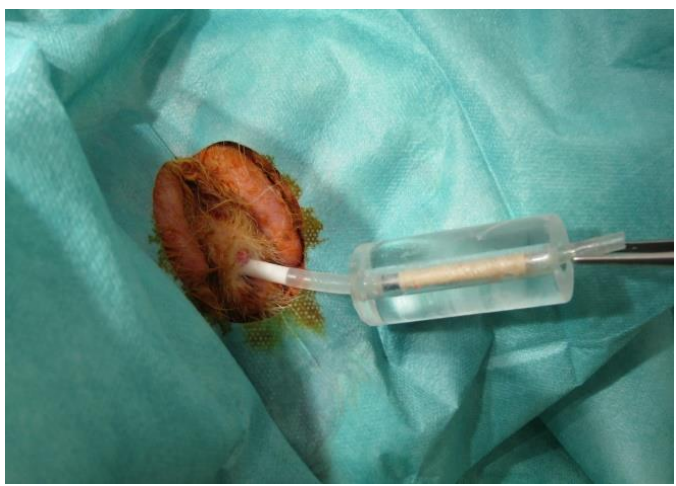

**Fig. 8:** Insertion of a $^{32}$P-foil catheter prototype inside a plexiglass radiation protector into a male rabbit urethra.

On day 35 after stenosis induction, thus 7 days after surgical treatment with catheter introduction and start of irradiation, the LDR brachytherapy was finished by removal of the catheter. After a further 4 weeks, on day 63 after beginning, the animals were sacrificed and their urethra prepared for later histopathological analysis focusing on early radiation induced effects. Urothelium inflammation was identified in tissue slides by influx of pseudoeosinophile granulocytes, which are a specific indicator for rabbits. An averaged inflammation grade by normalized influx of 0.83 for 0 Gy, 0.33 for 15 Gy and 1.17 for 30 Gy was determined, but without statistical significance. Mild radiation damage characterized by specific changes of medium and small vessels was observed for both dose groups. Although significant LDR-related biological effects could not be stated due to the short observation time and small number of animals, the intended feasibility and practicability of this treatment procedure could be demonstrated successfully.

## Discussion

Stenosis in hollow organs is a common clinical problem caused by a variety of reasons, often of iatrogenic traumatic or inflammatory etiology (Lumen 2009). Examples in urology include transurethral interventions or catheter inserts, including anastomosis strictures after surgical interventions such as prostatectomy, with an incidence rate of 1.1% to 8.4% (Elliott 2007). As a first therapy option, stenoses are usually dilated or slit with temporary catheterization.

Wound healing following this procedure, however, often leads to recurrence of stenosis and further operative revisions commonly end again in the same problem (Tritschler 2013). In this unsatisfactory situation, the proposed LDR brachytherapy approach could be an alternative possibility of stenosis or re-stenosis prevention.

For this reason, several studies based on brachytherapy have been performed using both commercially available systems and specific in-house developments of radiation sources. For comparison with the $^{32}$P-foil system described here, some typical irradiation devices should be mentioned as examples, which have already been used in clinical as well as preclinical studies on stenosis prophylaxis. In urology, after transurethral slitting of stenosis in 15 patients, a dose of 4 Gy in 3 mm tissue depth was applied directly after surgery on the following two or three days by inserting a commercial $^{192}$Ir afterloading line source through the lumen of the lying catheter (Olschewski 2003). Another clinical trial with $^{192}$Ir-HDR-Brachytherapy was conducted on 17 patients with different stricture history (Sun 2001). Here, a cumulative dose of 10 to 15 Gy was administered within 3 postoperative days. In animal studies, an experimental balloon catheter filled with the beta emitter $^{188}$Re ($t_{1/2}$ = 17 h, $E_{max}$ = 2.12 MeV) in liquid form, was used for HDR-brachytherapy (Shin 2006). After stenting of a dog urethra, doses of 20 and 40 Gy were delivered at 1 mm distance from the balloon surface adapting a technique, which has also been used for prevention of in-stent restenosis in coronary arteries. Another approach, somewhat similar to the present investigation, has been chosen in a preclinical LDR brachytherapy study on prevention of bile duct stenosis. Here, a self-expanding nitinol stent was covered with a PU film loaded with the beta emitter $^{166}$Ho ($t_{1/2}$ = 26.8 h, $E_{max}$ = 1.85 MeV) (Won 2005). Preliminary reports on prevention of bile duct as well as urethra stenosis with LDR brachytherapy using $^{32}$P-foils were presented earlier (Weidlich 2007, Assmann 2013).

The actual study presents an innovative and technically uncomplicated method, which may prevent surgical treated urethra from restenosis. The decision to rely on a radioactive foil has proved to be a good choice and fulfilled all requirements of the envisioned application: Foil production is a relatively simple and inexpensive process, a lot of biocompatible polymers are available, which can contain different compounds, and with PEEK even a rather radiation resistant polymer is at hand. In addition, PEEK with its high melting temperature enables different sterilization methods including gamma sterilization. Foils can be produced in large quantities, which can be stored, and activated and applied on demand. The dose distribution along a foil surface is more homogeneous as opposed to a nitinol stent, for example. The used 50 µm thick pre-rolled foils could be easily slid on the urinary catheter and fixed with shrink tubing, enabling an almost unchanged procedure of catheter implantation in clinics. The observed $^{32}$P wash-out may be further reduced by additional coating of the foil surfaces during production. Notwithstanding the simplicity of this approach, a foil-based implant is limited in its applicability and mainly suited for use in hollow organs. The implant production method, however, can be adapted to other applications. A new fabrication technique based on 3D printing (fused deposition modeling) of medical devices has become an emerging technology in recent years, which enables a much more flexible, yet personalized design of implants opening a new perspective for brachytherapy (Cunha 2020). Meanwhile, even PEEK based devices have been produced with this technique (Haleem 2019), which could incorporate also a $^{31}$P component, overcoming the geometrical constraint of foils and extending this therapy concept to further clinical challenges.

$^{31}$P is the only stable isotope of phosphor, thus during neutron activation no other eventually disturbing isotope besides $^{32}$P is formed. Selection of the nontoxic, temperature-stable and crystalline compound $NaPO_3$ was adequate, as it adds after activation only the short-lived

isotope $^{24}$Na to the radioactive inventory, which could even be used for calibration purposes. Considering the short range of $^{32}$P beta radiation, special care was taken to get the urethral tissue in close contact to the foil surface. Due to the surgical trauma an inflammation-related swelling in the stenosis area is to be expected, which additionally increases the tissue-foil contact, and thus ensuring, that the desired dose actually reaches the targeted volume: The proposed implant is in this sense self-adjusting. Assuming a homogeneous surrounding of the radiation source, calculation of dose distributions could be facilitated by using a fast dose kernel method. Thinkable extensions of this brachytherapy approach even to malignant diseases have to consider the beta radiation specific steep dose drop and the connected high contact dose, as shown in Fig. 3. For example, the irradiation of superficial dermatological tumors such as basalioma or melanoma with a "$^{32}$P film patch" could become an option, similar to the dura brachytherapy plaque used in Folkert et al (2015).

The urethra of male rabbits has shown to be a well-suited animal model for application tests of this new radioactive catheter. As a prerequisite for this study a reproducible stricture induction technique was essential. Using radial laser light irradiation visually guided by a pilot light, a circumferential stricture could be successfully and precisely located on all animals in the urethra. However, in order to get a significant therapeutic outcome with regard to brachytherapy, the time schedule of such a study needs to be correlated with the induced biological processes. As the mean acute wound healing processes takes 7-10 days, but complete healing success is finished after several months, it could be assumed that only a primary radiation effect will be observed during the defined follow-up time. The measured lowest inflammation grade within the 15 Gy group could be interpreted as indication of inflammation suppression at this dose. Interestingly, the overall mild radiation damage for up to 80 Gy contact dose correlates with the observed radiation tolerance of human urothelium (Olschewski 2003). Restenosis formation occurs mainly during the long healing process period, which was not in the scope of this animal experiment and needs further investigation.

The question of whether LDR or HDR brachytherapy is more suitable for prevention of postoperative stenosis or keloid scar formation and, more generally, the question whether there are different underlaying biological mechanisms of both irradiation modalities remains still open (Manimaran 2007). The presented preclinical LDR brachytherapy study using a $^{32}$P-foil catheter with its high reproducibility could motivate a comparative study with available $^{192}$Ir or $^{188}$Re HDR irradiation sources.

## Conclusion

Within this proof-of-concept study a novel approach for endoluminal LDR brachytherapy to treat restenosis in hollow organs has been developed and tested. The concept based on PEEK foils containing the beta emitter $^{32}$P has demonstrated to be a practicable and safe procedure during animal test on laser induced strictures of the rabbit urethra. This implant design, where the radioactive foil is wrapped around the urethral catheter, allows the application of a prescribed dose distribution in a precise and computable manner. A possible extension of this method to other clinical cases with similar restenosis problematics such as bile duct stenosis seems promising. As results of a dose response study in an animal model may considerably differ from human tissue response this part remains for further investigations within dedicated clinical investigations.

# Acknowledgments

This study was financially supported by the ''Bayerische Forschungsstiftung" (Projekt Betamod, 712/06). The authors thank Markus Bader, Michaela Püls, Henrike Otto, Kathrin Siegrist (Weick), Frank Strittmatter, Stephanie Uschold, Patrick Weidlich and Gabriele Wexel for their contributions. Special thanks also to Heinz Busch for his permanent support and Heinrich Seegenschmiedt for fruitful discussions.